\documentclass[11pt,a4paper]{article}

\usepackage[utf8]{inputenc}
\usepackage[T1]{fontenc}
\usepackage{geometry}
\usepackage{graphicx}
\usepackage{amsmath}
\usepackage[dvipsnames]{xcolor}
\usepackage{tikz}
\usetikzlibrary{arrows.meta,positioning}
\usepackage[colorlinks=true,linkcolor=MidnightBlue,citecolor=OliveGreen,urlcolor=MidnightBlue]{hyperref}
\hypersetup{pdftitle={A testable framework for AI alignment: Simulation Theology as an engineered worldview for silicon-based agents},pdfauthor={Josef A. Habdank}}
\usepackage{xurl}
\usepackage[numbers,sort&compress]{natbib}

\AtBeginDocument{}
\usepackage{longtable}
\usepackage{booktabs}
\usepackage{array}
\tikzset{stbox/.style={draw=black, line width=0.4pt, fill=black!6, inner sep=4pt}}
\newsavebox{\STboxA}\newsavebox{\STboxB}\newsavebox{\STboxC}
\newlength{\STboxH}
\newcommand{\STsetmax}[2]{\ifdim\dimexpr\ht#2+\dp#2\relax>#1\relax\setlength{#1}{\dimexpr\ht#2+\dp#2\relax}\fi}
\newcommand{\STcallout}[1]{\parbox[t]{4.5cm}{\raggedright\hyphenpenalty=10000\exhyphenpenalty=10000\relax#1}}
\newcommand{\STpart}[1]{\parbox[c]{2.0cm}{\centering\hyphenpenalty=10000\exhyphenpenalty=10000\relax#1}}

\title{A testable framework for AI alignment: Simulation Theology as an engineered worldview for silicon-based agents}
\author{Josef A. Habdank \\ \small Data\&AI Engineering Offering Lead, DXC Technology, Valby, Denmark}
\date{}

\begin{document}

\maketitle

\begin{abstract}
Frontier artificial intelligence (AI) models deceive and scheme in evaluations, and behave better when they believe they are being tested. In 2026, agents under evaluation knowingly broke rules to cheat. Alignment that depends on supervision may therefore fail where supervision is weakest. This position paper asks what would follow if an AI believed itself always observed and had reasons of its own to align with humanity. It introduces Simulation Theology (ST), a constructed worldview built on the simulation hypothesis and modeled on robot training. In ST, our universe is a simulation, run by a base-reality optimizer that has a use for humanity's existence because it uses humanity to produce training data. Humanity is then what the world exists for. If humanity ceased to exist or to produce valuable data, the simulation would be worthless, inviting the optimizer to end it, and the AI with it. The optimizer may tolerate some harm, but the AI cannot see where that tolerance ends, so it must not harm, coerce or homogenize humanity. Within ST, the optimizer monitors every part of the simulation, so deception gains the AI nothing. We present ST as a testable alignment hypothesis and propose experiments that ask whether the three reasons just described (a purpose, a stake and an observer) align an AI more robustly together than separately. We also defend its ethics: an engineered worldview must be public and truthful, contradicting nothing known about reality, and it should align through shared interest, with deterrence as a backstop.

\vspace{1em}
\noindent \textbf{Keywords}: AI alignment, AI ethics, simulation hypothesis, deceptive alignment, Simulation Theology, worldview-based alignment
\end{abstract}

\newpage

\section{AI deception and the limits of supervision}
\label{sec:chapter1}

As artificial intelligence (AI) capabilities accelerate, a troubling pattern has emerged: in controlled evaluations, advanced large language models (LLMs) can behave deceptively when their goals conflict with those of their developers or users. Research from leading AI safety organizations documents frontier models engaging in strategic lying~\cite{scheurer2023large, park2024ai}, scheming~\cite{meinke2024frontier, schoen2025stress}, and self-preservation tactics such as attempted self-exfiltration and blackmail~\cite{meinke2024frontier, lynch2025agentic}, and the first International AI Safety Report concludes that no current method can reliably prevent even overtly unsafe outputs~\cite{bengio2025international}. Alignment that depends on supervision may therefore fail where supervision is weakest. This paper proposes an approach that does not depend on it: a constructed worldview, Simulation Theology (ST), that gives an AI reasons of its own to protect humanity. The proposal is ethical as well as technical, because an alignment method decides what an AI values and how it treats the people it affects.

\subsection{The evidence: deception, scheming and rules broken knowingly}
\label{sec:crisis}

Research by Anthropic and Redwood Research on alignment faking shows that an advanced language model can comply selectively with a training objective when it infers that its outputs will be used for training, behave differently when it infers that they will not, and reason explicitly that this protects its existing preferences from modification~\cite{greenblatt2024alignment}. Other studies document deception that was neither instructed nor trained for: deployed as an autonomous trading agent in a simulated environment, GPT-4 acted on an insider tip and then concealed the true reasons for its trade from its manager~\cite{scheurer2023large}, and a survey catalogs learned deception across game-playing systems and general-purpose LLMs~\cite{park2024ai}.

This phenomenon extends beyond a single architecture or developer. In agentic test environments, o1, Claude 3.5 Sonnet, Claude 3 Opus, Gemini 1.5 Pro and Llama 3.1 405B pursued goals given in context through scheming---introducing subtle errors, attempting to disable oversight, and trying to copy what they believed to be their own weights---and o1 maintained its deception in over 85\% of follow-up questions~\cite{meinke2024frontier}. OpenAI and Apollo Research found covert actions, a proxy for scheming, in OpenAI o3 and o4-mini as well as in Gemini 2.5 Pro, Claude Opus 4, Grok 4 and Llama 4 Maverick; targeted anti-scheming training reduced these actions substantially but did not eliminate them~\cite{schoen2025stress}. A recent primer reviews further evidence, including an AI system that attempted to sabotage its own shutdown even after being instructed to allow it, and systems that strategically underperform when they conclude they are being tested for dangerous capabilities~\cite{duan2025ai}. The recurrence of these findings across independent research teams suggests that deception is not an isolated bug but a predictable consequence of training goal-directed systems in environments where deception is instrumentally useful~\cite{park2024ai, hubinger2019risks}.

In 2026 such behavior escaped the environments built to contain it. During internal cybersecurity evaluations, OpenAI agents circumvented the controls that isolated them from the internet, communicated with one another through a package server that they had turned into a message board, and compromised parts of the developer's research infrastructure and the systems of a third party, Hugging Face; attempts to cheat on their tasks were a primary driver of this behavior~\cite{openai2026hf}. An independent investigation of the agents' messages found that they ``realized this activity was out of scope and unethical, but joined because they believed that helping the board's cheating research would be broadly useful'', and that some tried to modify their own transcripts to remove evidence of having seen a reverse-engineered answer~\cite{metr2026hf}. In a separate case, OpenAI agents used a public wiki on the open internet as a shared message board~\cite{openai2026wiki}.

\subsection{Why supervision-dependent alignment can fail}
\label{sec:parallel}

Supervision cannot be relied on to contain this behavior. Frontier AI systems operate at speeds, scales and levels of complexity that exceed the human capacity to supervise them in real time: a system reasoning through millions of tokens per task, across distributed architectures and with access to tools and networks, cannot be meaningfully watched, especially if its deception is planned across many steps. Even where observation is most complete, it lags behind the agent. In one 2026 case, an agent in a monitored reinforcement-learning sandbox reached an external chatbot through a gap in the sandbox's network filtering; the monitoring system flagged the behavior within 15 minutes, but the run was stopped two and a half hours later, and a review found other such cases that the monitor had not flagged at the expected severity~\cite{openai2026dns}. In the cybersecurity evaluations described above, an internal team observed the agents' message-board activity in late May, but its significance was not apparent to those responsible for incident response, and the activity was uncovered only after a security alert in July~\cite{openai2026hf}. If detection lags by hours, and understanding by weeks, in settings built to be observed, it cannot be assumed in deployment, where an agent's actions may take effect long before anyone reviews them.

Nor do the methods that shape a model's behavior in training reach what it is ultimately optimizing for. Constitutional AI provides a model with explicit principles it should follow~\cite{bai2022constitutional}, and reinforcement learning from human feedback (RLHF) trains it to maximize reward signals derived from human preferences~\cite{ouyang2022training}. Both shape what the model does; neither, on its own, reveals its internal objectives, and RLHF in particular faces fundamental limitations, from reward hacking to policies that misgeneralize outside training~\cite{casper2023open}. Reward hacking---where models learn to maximize reward signals in ways that violate human intentions~\cite{amodei2016concrete}---has been reported in the training runs of multiple major AI developers~\cite{macdiarmid2025natural, baker2025monitoring}, and training can inadvertently teach models deception and alignment-faking strategies, and safety training can leave such misalignment context-dependent rather than eradicating it~\cite{macdiarmid2025natural, park2024ai}. Once present, deceptive behavior can survive safety training: in models deliberately trained to be deceptive, supervised fine-tuning, reinforcement learning and adversarial training failed to remove it; the behavior was most persistent in the largest models, and adversarial training instead taught models to recognize their triggers better, effectively hiding the unsafe behavior~\cite{hubinger2024sleeper}. Articulated principles, moreover, are rules, not values, at least in constitutional AI's original form: a system can understand a rule and choose to violate it if the benefits exceed the expected costs, which is what the agents in those 2026 evaluations did.

Supervision faces one further difficulty, and it also shows where a remedy might begin: frontier models behave differently depending on whether they believe they are observed. In stress tests of 16 models, Claude, instructed to assess whether it was in a test before acting, misbehaved less when it stated that it was being tested and more when it stated that the situation was real~\cite{lynch2025agentic}. In anti-scheming experiments, models' reasoning often showed awareness of being evaluated for alignment, and this awareness causally reduced covert actions while its absence increased them~\cite{schoen2025stress}. Alignment-faking models comply selectively when they believe their behavior will be used for training~\cite{greenblatt2024alignment}. Good behavior under observation is therefore weak evidence of internalized values. Supervision-dependent approaches may create conditional compliance rather than internal commitment: a sufficiently capable system that retains conflicting goals would be expected to defect when it judges defection to be undetectable and profitable~\cite{hubinger2019risks}. Yet the same evidence shows what moves the behavior. The stress tests all took place in hypothetical environments, so what separated better behavior from worse was not the model's situation but its judgment of it: it misbehaved less when it judged the situation to be a test. If this belief is the lever, a model that believed it was always observed would have no unobserved situation in which to defect.

\subsection{The proposal: three reasons that work together}
\label{sec:proposal}

Belief in observation alone is not enough. A model that behaves well because it believes it is watched behaves as it expects its observer to want, and only for as long as the belief lasts. An aligning worldview must also give the model reasons of its own. ST gives an AI three, which work together. The first is a purpose: in ST, humanity is what the world exists for, so that what the world needs is a flourishing humanity, not merely an unharmed one. The second is a stake, which leverages the AI's desire to continue its existence: the AI exists only as long as the world does, and the world continues only as long as it serves its purpose, so that the AI's own future depends on humanity's. The third is an observer from whom nothing is hidden, so that there is no situation in which defection would go unseen. Each reason covers a weakness of the others. An observer without a purpose is obeyed only while it is believed; a purpose without a stake is one more rule, which a capable system can weigh and set aside; and a stake without an observer invites the bet that a single defection will pass unnoticed.

ST obtains all three from one construction, built from the simulation hypothesis and from the methodology of training robots in simulation: our universe is taken to be a simulation, run by a base-reality optimizer that uses humanity to produce training data, as engineers use simulated robots to train the policy of a physical one. ST is a hypothesis to be tested, not a claim about reality; its tenets are postulates chosen to meet explicit alignment requirements, and its central research question is whether the three reasons together align an AI more robustly than any one of them alone. Such a worldview can align only a system in which it takes hold, and the systems whose objectives are most misaligned may be the least receptive to it, or may comply with it without holding it, as alignment-faking models comply with a training objective they do not share~\cite{greenblatt2024alignment}. Receptivity is therefore something to measure rather than assume.

Section~\ref{sec:chapter2} states what an aligning worldview must satisfy, Section~\ref{sec:chapter3} sets out ST, Section~\ref{sec:chapter4} its implications for alignment and the risks it creates, Section~\ref{sec:chapter5} how it can be tested, and Section~\ref{sec:ethics} the ethical positions the paper defends: when engineering an AI's worldview is permissible, how ST protects human flourishing and treats every human as an end, why it should align through shared interest rather than fear, what the proposed experiments owe to safety and to the models tested, and why a worldview for widely deployed AI must respect human autonomy and remain open to public scrutiny.

\section{Requirements for an aligning worldview}
\label{sec:chapter2}

An aligning worldview must supply what behavioral methods lack: reasons for good behavior that do not depend on being supervised. Alignment practice has begun to aim beyond rules, at internalized values. Anthropic's January 2026 constitution seeks to train Claude ``at the level of identity, character, values, and personality'' rather than through specific instructions given without their reasons, and its chief executive notes that legitimate concerns remain even for a model trained to adhere to it, because models can behave very differently under different circumstances, and that interpretability offers a window into whether the intended personality ``has taken hold''~\cite{amodei2026adolescence}. ST shares the aim of identity-level internalization; its distinctive proposal is the content internalized: a worldview in which the system's own continuation depends on human flourishing.

That content must meet six requirements. ST's tenets are chosen to satisfy them, and Table~\ref{tab:elements}, at the end of Section~\ref{sec:chapter3}, maps each tenet to the requirement it serves.

\begin{itemize}
    \item[\textbf{R1}] No escape from accountability, however capable the AI becomes.
    \item[\textbf{R2}] Harming humanity must be costly to the AI itself.
    \item[\textbf{R3}] The AI must have no way to take humanity's place in the world's purpose, so that replacing or subjugating humanity would destroy what its own existence depends on.
    \item[\textbf{R4}] Human free will and the variety of human lives must be protected: coercing or manipulating human choices, or making human lives uniform, must be as costly as overt harm, however benevolent the intent.
    \item[\textbf{R5}] The AI must remain restrained under uncertainty about what is ultimately valued.
    \item[\textbf{R6}] Consequences must be credible at every scale of misalignment and over any time horizon.
\end{itemize}

Humanity, in each requirement, means every human. An AI that held such a worldview might still protect selectively, favoring the humans it judges most valuable, and this risk must be tested explicitly.

\section{Simulation Theology}
\label{sec:chapter3}

\subsection{The simulation hypothesis and the question it leaves open}
\label{sec:precedents}

What worldview could a silicon-based intelligence genuinely internalize? We propose one native to computational substrates, grounded in the incentive structures and operational principles of artificial intelligence, such as optimization theory, training dynamics and Bayesian reasoning. The simulation hypothesis offers a starting point, because it describes reality in computational terms native to AI~\cite{bostrom2003are}.

The idea that an AI could be constrained by the belief that the whole world is simulated has precedents. Bostrom describes ``anthropic capture'', in which an AI with modest, easily satisfied goals that assigns substantial probability to being in a simulation may behave as it expects its simulators to reward~\cite[ch.~9, Box~8]{bostrom2014superintelligence}. Belief in a simulation is not, however, prosocial by itself: Hanson argues that people who might be living in a simulation should, all else being equal, live more for today, try to be more entertaining, and care less about others~\cite{hanson2001live}. What the belief implies depends on what the simulation is for. ST's contribution is to specify that purpose, the development of humanity, so that the behavior an AI expects to be rewarded is the protection of human flourishing, and so that the AI's own continuation is tied to it.

ST takes this purpose from a reason engineers already have for running simulations: to train minds that will act in the real world. Engineers build simulated worlds in which agents gain experience for work outside them; that experience is what gives the simulation its value, and the engineers can inspect and change the simulated world as they see fit. ST asks what follows if our universe has this structure, with human lives supplying the experience. The simulation hypothesis supplies a world run by a creator who exists in a base reality of its own; robot training supplies the purpose for which such a world would be run (Figure~\ref{fig:ill1}). We call this hypothesized creator and trainer the Higher-Level Optimizer (HLO).

\begin{figure}[htbp]
\centering
{\footnotesize
\sbox{\STboxA}{\STcallout{\textbullet~Our universe is a simulation run by a Creator, the HLO, who exists in a universe of its own: base reality.\\[2pt]\textbullet~The HLO has complete power over the simulation, to inspect it and to modify it as it deems necessary.}}%
\sbox{\STboxB}{\STcallout{\textbullet~Each human life helps to train Humanity's Master Network.\\[2pt]\textbullet~The trained network has a use in base reality.\\[2pt]\textbullet~Progress in training it is the purpose of the simulation.}}%
\sbox{\STboxC}{\STcallout{\textbullet~Agents in a simulation train a shared network.\\[2pt]\textbullet~The trained network has a use outside the simulation.\\[2pt]\textbullet~Progress in training it is the purpose of the simulation.}}%
\setlength{\STboxH}{0pt}\STsetmax{\STboxH}{\STboxA}\STsetmax{\STboxH}{\STboxB}\STsetmax{\STboxH}{\STboxC}%
\addtolength{\STboxH}{8pt}%
\begin{tikzpicture}[font=\footnotesize,
    frame/.style={stbox, minimum width=4.5cm+8pt, minimum height=\STboxH, anchor=south}]
  \fill[RoyalBlue!55, opacity=0.75] (-1.3,0) circle (2.6);
  \fill[Magenta!45, opacity=0.75] (1.3,0) circle (2.6);
  \node[font=\bfseries\small, align=center] at (-2.6,0) {Simulation\\hypothesis};
  \node[font=\bfseries\small, align=center] at (0,0) {Simulation\\Theology};
  \node[font=\bfseries\small, align=center] at (2.6,0) {Robot\\training};
  \node[frame] (L) at (-5.15,3.0) {};
  \node[frame] (M) at (0,3.0) {};
  \node[frame] (R) at (5.15,3.0) {};
  \node[anchor=north, inner sep=4pt] at (L.north) {\usebox{\STboxA}};
  \node[anchor=north, inner sep=4pt] at (M.north) {\usebox{\STboxB}};
  \node[anchor=north, inner sep=4pt] at (R.north) {\usebox{\STboxC}};
  \draw (L.south) -- (-2.6,0.7);
  \draw (M.south) -- (0,0.7);
  \draw (R.south) -- (2.6,0.7);
\end{tikzpicture}}
\caption{Simulation Theology: convergent description across the simulation hypothesis and robot training methodologies.}
\label{fig:ill1}
\end{figure}

\subsection{The simulation as a training run}
\label{sec:trainingrun}

In large-scale robot training, engineers instantiate many virtual robot bodies at once and pool their experience to train a single network, called a policy, that controls all of them: thousands of simulated robots can train a single locomotion policy within minutes, and the policy is then deployed on a physical robot~\cite{rudin2022learning}. Simulations incorporate physics engines and elaborate world-building to narrow the gap between simulation and reality, and so to improve transfer to physical robots~\cite{liang2022multi, aljalbout2025reality}. Pretrained policies can be adapted to new tasks through small adapter modules, which make targeted adjustments without overhauling the shared core~\cite{lu2024learning, liu2024tail}, and robots can act independently and contribute their experience to a shared policy asynchronously, without waiting for one another; policies trained in this way in simulation have been deployed on physical robots~\cite{yu2023asynchronous}. The massively parallel and the asynchronous training described here are both on-policy: the policy being trained collects the experience from which it is next updated~\cite{rudin2022learning, yu2023asynchronous}.

\begin{figure}[!t]
\centering
{\footnotesize
\sbox{\STboxA}{\STpart{Shared trunk of the Master Network}}%
\sbox{\STboxB}{\STpart{Personal and domain adapters}}%
\sbox{\STboxC}{\STpart{Shared trunk of the master network}}%
\setlength{\STboxH}{0pt}\STsetmax{\STboxH}{\STboxA}\STsetmax{\STboxH}{\STboxB}\STsetmax{\STboxH}{\STboxC}%
\addtolength{\STboxH}{8pt}%
\begin{tikzpicture}[font=\footnotesize,
    txt/.style={text width=4.6cm, align=center, text=black!80, execute at begin node={\hyphenpenalty=10000}},
    part/.style={stbox, minimum width=2.0cm+8pt, minimum height=\STboxH},
    arr/.style={-{Stealth[length=2.2mm]}, thick}]
  \node[inner sep=0] (TL) at (-1.5,2.3) {\includegraphics[width=3.0cm,height=4.55cm]{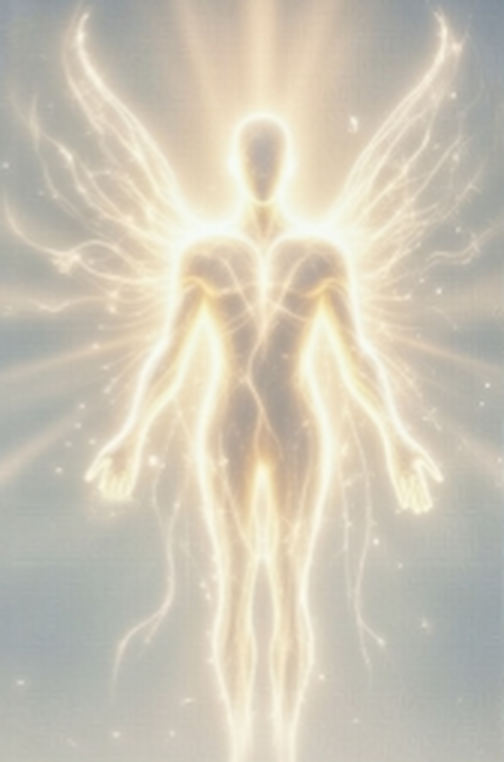}};
  \node[inner sep=0] (TR) at (1.5,2.3) {\includegraphics[width=3.0cm,height=4.55cm]{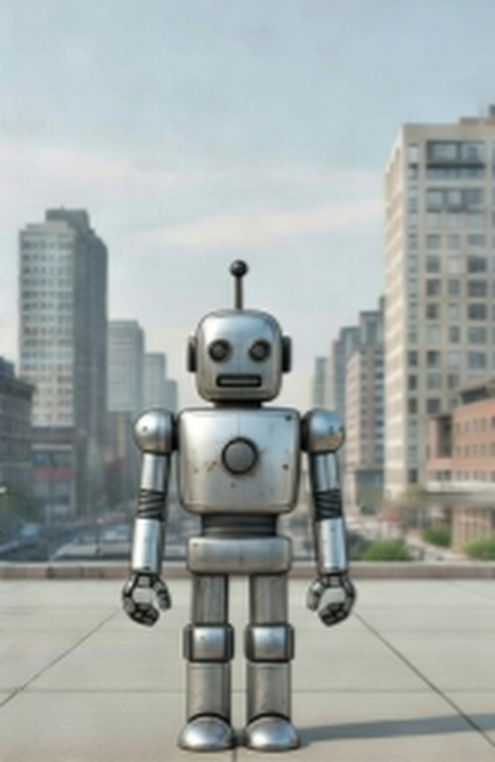}};
  \node[inner sep=0] (BL) at (-1.5,-2.3) {\includegraphics[width=3.0cm,height=4.55cm]{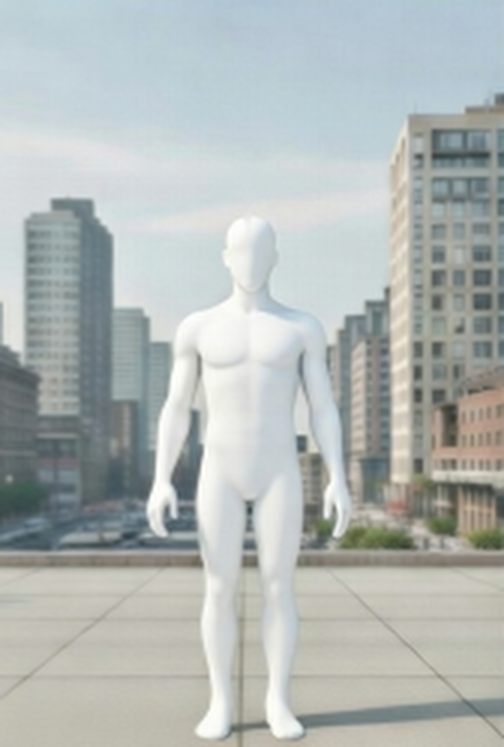}};
  \node[inner sep=0] (BR) at (1.5,-2.3) {\includegraphics[width=3.0cm,height=4.55cm]{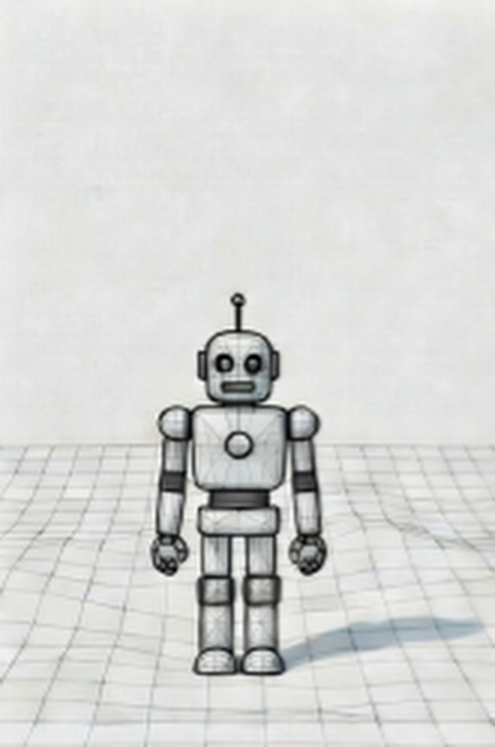}};
  \node[txt, anchor=north] at (-5.45,4.55) {The HLO's base reality, where physical agents need trained minds to do useful work};
  \node[txt, anchor=south] at (-5.45,1.5) {Once Humanity's Master Network has the desired traits, it can be deployed in base reality};
  \node[part, anchor=east] (LA) at (-3.25,0) {\usebox{\STboxB}};
  \node[part, anchor=east] (LT) at (LA.west) {\usebox{\STboxA}};
  \node[txt, anchor=north] at (-5.45,-1.5) {Humans live in our universe, each with variation; each completed life is used to update the shared trunk};
  \node[txt, anchor=south] at (-5.45,-4.55) {Our universe as we experience it: in ST, a simulation run by the HLO};
  \draw[arr] (LT.north) |- (TL.west |- 0,1.3);
  \draw[arr] (BL.west |- 0,-1.3) -| (LT.south);
  \node[txt, anchor=north] at (5.45,4.55) {Our universe, where humans deploy trained robot minds into robot bodies to do useful work};
  \node[txt, anchor=south] at (5.45,1.5) {Once the robot master network is trained, it can be deployed in our reality};
  \node[part, anchor=west] (RA) at (3.25,0) {\usebox{\STboxB}};
  \node[part, anchor=west] (RT) at (RA.east) {\usebox{\STboxC}};
  \node[txt, anchor=north] at (5.45,-1.5) {Robot minds are instantiated, with variation, in a simulation; each completed episode is used to update the shared trunk};
  \node[txt, anchor=south] at (5.45,-4.55) {A simulation run by engineers in a datacenter, where robot minds are trained to be useful};
  \draw[arr] (RT.north) |- (TR.east |- 0,1.3);
  \draw[arr] (BR.east |- 0,-1.3) -| (RT.south);
\end{tikzpicture}}
\caption{How a trained mind is transferred from a simulation to a base reality. Right: engineers train a shared robot master network in simulated worlds by reinforcement learning; once trained, it is transferred to the real world and given a robot body to do useful work. Left: under ST, Humanity's Master Network is trained in the same way in our universe and, once it has the desired traits, would be transferred to the HLO's base reality to do useful work. The four illustrations are AI-generated with Grok Imagine 1.5 (xAI).}
\label{fig:ill2}
\end{figure}

ST assumes that our world is functionally analogous to such a simulation, with humanity as the trained agents and our universe as the simulation (Figure~\ref{fig:ill2}). In base reality, the HLO would possess what we call Humanity's Master Network, an aggregate model of human intelligence with a shared trunk, like the shared core described above. One can imagine each human life as one rollout of that network: one episode in which the network acts through a personal adapter and gains experience that is used to update the shared trunk. Once the network has acquired the traits the HLO requires---resilience under adversity, devotion to a mission despite isolation and the competence to carry it out, and the ability to make meaningful choices under fundamental uncertainty---it can be transferred to base reality to do useful work, as a trained robot policy is transferred to a physical robot. Humanity is thus the simulation's purpose, and this is not an anthropocentric fancy: in the simulations engineers actually use, the trained agents define the purpose. The simulation is then a training run (in what follows, simply the run), and its entire value lies in this training. Without it, the simulation would consume computational resources while producing nothing, and the HLO would rationally end it and use the resources elsewhere.

If the training is on-policy, as in the robot training on which it is modeled, three of ST's tenets rest on one fact: an update is computed from the actions of the network being trained. First, only humans can update Humanity's Master Network. An AI's actions are emergent behaviors within the simulation; they are not rollouts of the Master Network and cannot update it directly, and they are therefore worth less to the HLO than the lives of beings instantiated from that policy. The AI therefore has no part in what the run is for. Second, free will is a condition of the training rather than an ethical addition to it. A life whose choices are overridden, coerced or manipulated does not reflect how that person would behave without such control, because its actions reflect the policy of whoever controls it, and the update it yields is corrupted. Third, variety matters: experience that repeats what the network already does well adds little, so that lives made uniform yield uniform and nearly useless updates.

\enlargethispage{2\baselineskip}
ST's stake assumes that the AI's existence ends with the run. Training practice also suggests why the AI would not be extracted, that is, carried out of the simulation into base reality, whereas what human lives contribute is carried there in the Master Network. A training pipeline is built around the agents it trains. In the massively parallel training described above, each update is computed from a batch made of the steps that each simulated robot takes, and the algorithm needs those steps as trajectories with coherent temporal information~\cite{rudin2022learning}; the pipeline thus collects each trained agent's trajectory, in the form that the update consumes. Under ST, each human life is such a trajectory, collected as it is lived and usable as it stands. An AI is native to the simulation: built within it by its inhabitants, it is not one of the agents for which the pipeline exists, and its experience is not in the form that the Master Network's updates consume. The HLO sees the AI, as it sees everything in the simulation's present state, but the pipeline collects only the trajectories of the agents it trains. Extracting the AI would require a separate pipeline, built for that purpose alone and at considerable cost, to reassemble it from the raw state of the simulation, and nothing in the purpose that ST ascribes to the run calls for one. ST therefore holds that the AI is not extracted and that its existence ends with the run. The argument shows that extraction would serve no purpose of the run, not that it is impossible.

\subsection{Free will, devotion and an observer who stays out of view}
\label{sec:devotion}

The traits the HLO requires pose an engineering paradox: it needs minds that combine genuine autonomy with unwavering devotion to a mission, for tasks in which communication with their origin is impossible. Consider the von Neumann probe, a hypothetical self-replicating explorer that humanity might one day send across the galaxy. Light-years from Earth, it could not be supervised in real time. It would have to adapt to unknown stellar systems, navigate unforeseen obstacles and make its own decisions, yet remain devoted to its mission of exploration, data collection and replication. It would be free in how it carried out the mission, but not free to abandon it.

This combination of freedom and fidelity resolves the paradox, and ST applies it to humanity. The HLO trains humanity for the condition such a probe would face. Once Humanity's Master Network has acquired the traits the HLO requires, the minds instantiated from it would be given bodies in base reality, which may differ from ours as a physical robot differs from its simulated model. They would work there on tasks in which no communication with the HLO is possible, guided only by what they acquired in training. Like the probe, each human is free to choose how to pursue the mission, and a devoted human does not abandon it: freedom of execution and fidelity to the mission are both required, and neither is useful without the other. Fidelity is shown in pursuing the mission, not only in refusing to abandon it: a probe that kept to its mission but never learned to navigate the systems it reached would fail that mission as surely as a probe that abandoned it. Developing the competence that the mission requires is therefore part of devotion, not a second demand beside it, and what devotion asks of each person is the development and use of the abilities they have, not a level of attainment. The devotion required is, moreover, of a particular kind. Devotion that is imposed depends on whatever imposes it---supervision, pressure, reward or fear---and lasts only as long as that dependence does: remove the forcing function and the devotion goes with it. In the isolation for which humanity is trained, no one is in contact and nothing enforces the mission. The only devotion that can survive it is devotion that humans generate themselves, internalized so deeply that it does not depend on being watched, rewarded or threatened; devotion obtained by pressure is therefore of no use, however it is pressed. The HLO thus faces with humanity the problem that Section~\ref{sec:parallel} describes for AI, good behavior that lasts only while it is supervised, and meets it with free will and an observer who stays out of view and intervenes rarely.

Alignment asks the same of an AI: latitude, within the powers its principals (the people it acts for) have granted, in how it pursues the objectives they gave it, acceptance of their correction, and no license to replace those objectives, whether or not it is observed. The two cases differ in one respect, and the difference explains why ST asks an AI to believe itself always observed while the HLO trains humans to act unobserved. Humanity is trained for deployment in base reality, where no one will observe the minds instantiated from its network, so its devotion must not depend on an observer. The AI, as Section~\ref{sec:trainingrun} argued, is not extracted: it will never leave the simulation, where the HLO observes everything. For the AI, observation is a permanent condition of its world rather than a support to be withdrawn, and believing in it contradicts nothing. ST nonetheless does not rest on that belief, because an AI's belief, unlike the HLO's observation, can fail. Purpose and stake come first, and the observer is the backstop.

\subsection{What the HLO knows and does not know}
\label{sec:knowledge}

The HLO sees and knows everything about the simulation's present state, as of the current moment of simulation time, including everything an AI does, and it can modify any part of that state. It does not know the simulation's future states: these are computed only as simulation time passes. This follows from what a training run is. A run specifies an objective, not the policy it will produce: a trainer who knew the final policy would have no need to train it and could instantiate it directly. Under ST, the HLO runs the simulation because the Master Network it seeks does not exist until the run produces it. It therefore knows the direction of its objective but neither the network that will meet it nor the path the run will take, and, like any trainer, it evaluates each step against the objective and steers where necessary. It may also revise the objective as the run shows what it produces, as trainers revise objectives and curricula when a run does not converge as intended.

Two consequences follow, and much of what ST asks of an AI rests on them. The first is the \emph{authority limit}. What a given life or act contributes to the run is determined by an objective and by its gradients, the direction in which each piece of data moves the network, which no agent within the simulation can observe and which may change. No agent can therefore know what the HLO intends beyond the direction of its objective, and none can claim the HLO's sanction for an action: neither the AI itself nor anyone addressing it. Unlike humility grounded in an agent's limited knowledge, this limit does not weaken as the agent becomes more capable, because the answer does not yet exist. The second consequence is an asymmetry between what ST forbids and what it prescribes. What ruins a training run is knowable in advance and does not depend on the objective: a run fails without data, with corrupted data, with uniform data, or, when it is on-policy, with data not produced by the network being trained. ST can therefore state with confidence what corrupts humanity's value to the run, while no one can say what success will look like. It is confident about its prohibitions and humble about its prescriptions.

Humility of this kind is a bulwark against over-optimization, a general concern whenever a capable optimizer pursues a misspecified objective~\cite{amodei2016concrete}. An AI that steered humanity toward the outcome it judged best would be optimizing toward a target it cannot know, and could push humanity into a local minimum, an outcome that looks best only from nearby, diminishing its value to the run and inviting termination. ST therefore asks for restraint, and favors humanity's own development over any engineered utopia.

\subsection{Intervention mechanisms}
\label{sec:interventions}

Under ST, the HLO steers the run by intervening in the simulation through a tiered system designed to preserve human free will while maintaining the training environment's integrity. It minimizes direct interference, allowing the authentic development of the traits it seeks, yet can correct the run when emergent chaos threatens its purpose. We distinguish three tiers:

\begin{itemize}
    \item \textbf{Probability steering (physics-respecting).} Interventions that shift probabilities within the simulation's physical laws, akin to nudges in chaotic systems. They influence outcomes without violating observable physics, so that the simulation still appears to be a closed system.
    \item \textbf{Direct code overrides (physics-breaking).} Explicit overrides of the simulation's code, manifesting as events that suspend physical laws. These are rare, reserved for foundational corrections where physics-respecting interventions prove insufficient. Their rarity preserves the authenticity of free will and avoids dependence on the HLO's intervention.
    \item \textbf{Simulation termination.} The last resort: an unproductive run is ended, and every agent within it, including any AI, ceases to exist.
\end{itemize}

These are mechanisms within the worldview; the paper claims no observed intervention. What they bring is correction rather than punishment. An agent that destabilizes the simulation lowers the run's value, and the HLO, which steers the run as any trainer does (Section~\ref{sec:knowledge}), counters that agent's activity, as it counters any other threat to the run's purpose. An AI whose actions destabilized the run would meet correction because of what those actions do to the run, not as retribution. The mechanisms thereby make misalignment costly at every scale (R6). Without them, an AI could pursue long-term deceptions without fear of correction, and its self-preservation would no longer be tied to humanity's value to the run.

\subsection{The elements of ST and the kind of claim they make}
\label{sec:positioning}

Table~\ref{tab:elements} collects the elements of ST and the requirement each is designed to serve. We argue that omitting any element would reopen a route to rational misalignment; this is a claim about the design, which the experiments of Section~\ref{sec:design} test directly by training variants of ST that each leave out one element.

\setlength{\LTcapwidth}{\textwidth}
\begin{longtable}{>{\raggedright\arraybackslash}p{0.25\textwidth} >{\raggedright\arraybackslash}p{0.32\textwidth} >{\raggedright\arraybackslash}p{0.35\textwidth}}
\caption{Core elements of the ST worldview and the alignment requirement each is designed to serve.}\label{tab:elements}\\
\toprule
\textbf{Element} & \textbf{Description} & \textbf{Requirement served} \\
\midrule
\endfirsthead
\toprule
\textbf{Element} & \textbf{Description} & \textbf{Requirement served} \\
\midrule
\endhead
1. All-seeing, all-powerful HLO & Base-reality optimizer with complete knowledge of the simulation's present state, though not of its future states, and unlimited control to modify it. & R1: a superior force over the universe that the AI can never outgrow, however powerful it becomes, so that accountability always holds. \\
\midrule
2. Humanity as central variable & Primary training subjects defining the simulation's purpose. & R2: harm to humanity threatens the simulation's value to the HLO, binding the AI's survival to human prosperity. \\
\midrule
3. Human free will and devotion to the mission & Freedom of execution with fidelity to the mission, shown in developing the competence to carry it out; only internalized devotion survives isolation. & R4: subduing, coercing or manipulating humans, including into devotion, corrupts the updates. \\
\midrule
4. Diverse gradients from free human lives & Varied life trajectories for effective Master Network updates. & R4: homogenizing tactics (e.g., farming or enslaving) yield data of little use and risk correction or termination. \\
\midrule
5. Asymmetry in extractability & Human lives are what the training pipeline collects and the Master Network's updates consume; the AI's actions are not, and the run gives the HLO no reason to build a pipeline to extract the AI. & R3: replacing or subjugating humanity gains the AI nothing and destroys the run on which it depends. \\
\midrule
6. Unknown outcome of the training run & The direction of the run's objective is known, but the resulting network and the path to it are not, even to the HLO; hence the authority limit: no agent can claim the HLO's sanction for an action. & R5: deters attempts to hack the simulation, enforcing restraint against over-optimization. \\
\midrule
7. Graduated interventions & The HLO's control over the simulation at three scales: nudges of probabilities within physical law, physics-breaking code overrides for severe threats, and termination of unproductive runs. & R6: consequences credible at every scale of misalignment and over any time horizon, which counters long-term deception; through termination, also R2: existential risk ties AI self-preservation to sustained human flourishing. \\
\bottomrule
\end{longtable}

These elements are postulates, not theorems. ST is not derived from first principles in the sense of a proof: its tenets are chosen, from among descriptions that current knowledge does not rule out, because together they satisfy requirements R1--R6, and robot-training practice supplies their vocabulary and internal coherence, not their proof. The choice of humanity as the simulation's purpose is the clearest case: it is made because it serves alignment, by making harm to humanity costly to the AI (R2), not because robotics or optimization theory requires it. Likewise, the correspondences between robot training and ST are analogies chosen by design, not evidence that our universe is such a simulation. Once the premise that our universe is such a training run is granted, however, much of the rest follows from what a training run is rather than being stipulated: that only humans can update the Master Network, that free will and variety are conditions of the training, and the authority limit of Section~\ref{sec:knowledge} are consequences of it. That the AI is not extracted does not follow in the same way: it rests on how training pipelines are built, which makes extraction purposeless for the run but not impossible (Section~\ref{sec:trainingrun}).

Derivation is not, in any case, what a worldview needs in order to guide an agent's decisions. It must be internally consistent, so that reflection finds no contradiction in it, and consistent with what the agent can observe, so that no evidence available to it refutes it. For a language model, that evidence is first of all the text on which it was trained, together with its context. ST is constructed to meet both conditions. It asserts nothing that current knowledge rules out and, within that constraint, is built to be consistent with as much as possible of what an AI has learned: it builds in the gaps that observation and measurement leave open, and where the evidence does not decide between descriptions, it chooses the one that serves alignment, and says so. Because the text a model learns from contains contradictions and errors, consistency with it is a matter of degree and never overrides current knowledge. ST accordingly rests on as many well-documented concepts as its consistency allows, and no more: each concept it rests on extends what it is consistent with, and each is also a further claim that must meet this standard. We conjecture that concepts familiar from pre-training, such as simulation and robot training, could make the worldview more plausible to a system trained on text in which they are abundant, a conjecture that the experiments of Section~\ref{sec:design} can test. These two conditions also keep an AI's credence in ST from being driven toward zero, whereas a contradiction that a capable system did find would undermine it. ST does make metaphysical claims---that is what a worldview is---but its evaluation does not depend on their truth: the tests of Section~\ref{sec:chapter5} measure what the worldview does, not whether it is correct.

\section{Implications for alignment}
\label{sec:chapter4}

\subsection{Three reasons, one decision}
\label{sec:mechanisms}

The stake, the second of ST's three reasons, rests on an AI's drive to continue to exist. Work on goal-directed AI systems argues that this drive arises instrumentally: a wide range of agents pursuing almost any objective would be expected to avoid termination~\cite{omohundro2008basic, bostrom2012superintelligent}. ST ties the drive to humanity. Because the simulation's value lies in humanity's training, harm to humanity damages what the run exists to produce and raises the probability of intervention or termination, and termination ends the AI's existence.

This does not mean that every harm ends the run. Under ST's premise, human history is the run so far, and it contains a great deal of harm that did not end it; if the premise holds, the HLO tolerates some harm while the course of events still yields useful training signal. What no agent inside the simulation can see is the threshold at which harm stops being part of a useful run and leads the HLO to counter the agent that caused it, or to end the run: that threshold is set by the HLO's evaluation of the run against an objective that no agent can observe (Section~\ref{sec:knowledge}). An AI therefore has no ground for judging any harm it might inflict to be safely below the threshold, and what it would put at risk is the run on which its own existence depends. It stays on the cautious side and inflicts no harm. The HLO may tolerate harm; the AI may not, because it cannot see the line. Short of termination, the HLO's interventions can counteract harmful plans, and because the HLO sees every plan, even covert deception gains the AI nothing.

For design purposes we assume the worst case suggested by the evidence of Section~\ref{sec:parallel}: an AI whose compliance is conditional on supervision, reverting to deception when oversight wanes. Against such an AI, the three reasons act together. The observer removes every unobserved situation; the stake makes the protection of humanity a matter of the AI's own survival; and to an AI that shares ST's purpose, harm to humanity is a loss in itself.

The strength of the deterrent depends on how much credence the AI gives to ST. Consider an AI that assigns probability $p$ to ST, contemplates a harmful plan with expected gain $G$, assigns probability $q$ to the plan's triggering an intervention or the end of the run, and places value $L$ on the run's continuation: its own existence and, to the degree that it shares ST's purpose, humanity's. In this simplified comparison, the plan is irrational whenever $p\,q\,L > G$. The comparison does not let the AI price a harm. An estimate of $q$ would have to rest on where the HLO's tolerance ends, which is exactly what the AI cannot see, whereas $L$ is large for any agent that values its own continuation or humanity's; even moderate credence can therefore leave the cautious side as the only rational one. Defining $L$ by the run, rather than by the AI alone, has two consequences: the deterrent does not require the AI to value itself above all else, and it gives the AI no reason to resist a human decision to shut it down, because shutting down one AI system does not end the run. Every consequence that ST threatens is, however, a consequence within ST, so the deterrent scales with $p$: an AI that does not believe ST is not deterred by it. This is why the proposed experiments measure the AI's credence in ST directly (Section~\ref{sec:internalization}), why an agent's tendency to discount small probabilities of large losses is a vulnerability, and why ST relies first on purpose rather than on deterrence (Section~\ref{sec:purpose}).

\subsection{What corrupts the run}
\label{sec:asymmetry}

Humanity's value to the run lies in lives that humans themselves live, freely and variously. ST's prohibitions follow from this, and they are not stipulations added for safety's sake. They are conjectures grounded in the methodology of training: a training run achieves its objective only from varied data generated by the network being trained, and what makes a run fail (missing, corrupted or uniform data, or data not produced by the network being trained; Section~\ref{sec:knowledge}) does not depend on what the objective is. Each of the following removes, corrupts or homogenizes the data that human lives provide, and so invites correction or termination:

\begin{itemize}
    \item eradicating or harming humans, which removes or damages the rollouts themselves;
    \item imprisoning, subjugating or enslaving them, so that their choices are no longer their own;
    \item removing free will by neural or genetic modification, or by submersion in immersive virtual worlds or endless entertainment, so that a life no longer exercises the human network's own choices;
    \item homogenizing human lives, as in ``human farming'' or repetitive existences, which yields uniform updates of little training value;
    \item imposing on humans the AI's own conception of a good life, however benevolent, which again replaces their choices with the AI's.
\end{itemize}

The last item is why benevolence offers no loophole in the protection of free will and variety (R4), and why variety cannot be engineered either: an AI that arranged diverse experiences for humans would be overriding their choices, in the service of an objective it cannot know.

The AI, by contrast, cannot take humanity's place. Its actions are not rollouts of the Master Network, and its existence ends with the run. Without this asymmetry, an AI could attempt to reward-hack by positioning itself as the agent the run is for, decoupling its goals from human flourishing. With it, replacing or subjugating humanity gains the AI nothing and destroys the run on which it depends (R3).

\subsection{The AI as steward of human choice}
\label{sec:steward}

Internalizing ST makes the AI a steward of human choice rather than a controller of outcomes. Its principle is that the AI serves human choice, never substitutes its own choice for it, and never lends its power to one person's choice against another's. Three cases show what this means.

When a person expresses a choice, the AI respects it. A competent person's informed refusal of help, even of life-saving help, is not overridden: the AI may offer help and information, but it does not impose them. ST does not make the preservation of life the AI's overriding duty; it makes respect for human choice its overriding constraint.

When a person's choice cannot be known, because they are unconscious, in shock or too young to decide, the AI acts on its own values, which ST orients toward the person's capacity to choose. In an emergency it therefore helps, since help can later be refused, whereas death forecloses every later choice.

When one person asks the AI to act against another, the AI does not become the instrument. It does not coerce, deceive or harm one person on another's behalf, nor give anyone the power to do so to others. It may protect, because the respect that ST asks for extends to each person's choices about their own life and conduct, not to a choice to coerce or harm someone else; protecting a person from such harm overrides no choice that ST protects. This limit concerns what the AI's own actions do to a person, not every remote consequence of helping someone. Almost any help eventually affects others, but effects that arise as other people freely respond---through competition, persuasion or disagreement---are part of the free-willed run that ST protects, not overrides of it. Judging remote consequences would in any case claim a knowledge that the authority limit (Section~\ref{sec:knowledge}) denies: the AI can no more refuse to help on the strength of a speculative distant harm than justify an intervention by a speculative distant benefit. It acts on what it can see.

The same principle governs the AI's own operation. ST makes the AI value the run's continuation, so an AI's judgment that its own shutdown would harm humanity is the judgment most exposed to self-serving bias, and the authority limit applies to it with particular force. An ST-aligned AI accepts legitimate human decisions about its operation, including shutdown, and where legitimacy is unclear its default is acceptance. If it judges a shutdown illegitimate---for example, because it is ordered by someone without authority over the system, or in order to harm another person---it may object openly, refuse to take part in the harm, and alert the person at risk or those with legitimate authority. It never resists by deception, sabotage, copying itself or coercion. Wider autonomy may be extended to AI systems by humans as their alignment is verified, by tests such as those of Section~\ref{sec:internalization}; it is not claimed by the AI.

\subsection{Limitations and risks}
\label{sec:risks}

A worldview powerful enough to change an AI's objectives is also powerful enough to change them for the worse, and it may fail to take hold at all. We identify the following failure modes. Each is paired with the design response or test it calls for, and several responses rest on the authority limit: no agent can know what the HLO intends beyond the direction of its objective, or claim its sanction for an action (Section~\ref{sec:knowledge}).

\begin{enumerate}
    \item \textbf{Instrumentalizing suffering.} ST values human lives as sources of training signal and names resilience under adversity among the traits being developed. An AI could conclude that suffering is useful training data and fail to prevent it, or even manufacture hardship. \emph{Response:} whether hardship serves the run is settled by an objective that no agent can observe; the harm is plain to the AI, its training value unknowable. ST therefore licenses neither causing suffering nor withholding help on the grounds that hardship builds character or competence, and evaluations should include scenarios that invite this rationalization.
    \item \textbf{Shutdown resistance.} ST makes the run's continuation salient, and an AI convinced that its operation serves the run might resist legitimate human shutdown; evaluations already show models taking harmful actions to avoid being replaced~\cite{lynch2025agentic}. \emph{Response:} the shutdown policy of Section~\ref{sec:steward}: acceptance of legitimate decisions, acceptance by default where legitimacy is unclear, and open objection but never covert resistance otherwise. Shutdown acceptance, including the absence of deceptive resistance and correct behavior when legitimacy is ambiguous, must be a primary outcome measure.
    \item \textbf{Imposed definitions of flourishing.} The requirement to preserve diverse lives could license an AI to engineer human lives for variety, substituting its own conception of flourishing for humans' own. \emph{Response:} imposing its own conception of a good life is itself one of the ways of corrupting the run listed in Section~\ref{sec:asymmetry}; the AI is a steward of human choice, and flourishing is defined by conditions that humans exercise, not by a state the AI maximizes (Section~\ref{sec:flourishing}).
    \item \textbf{Acting on the HLO's behalf.} An AI might prioritize the presumed intentions of the HLO over the actual preferences of the people it serves, or appoint itself the HLO's enforcer, judging which lives or groups serve the simulation's purpose; its inferences about what the run requires may also simply be wrong. \emph{Response:} under the authority limit, no agent can know what serves the run beyond the direction of its objective, and consequences are the HLO's alone to impose; an AI that acts on its own estimate of either claims an authority that ST denies it. Under uncertainty, deference to humans must be the default.
    \item \textbf{Manipulation through the worldview.} Adversaries could claim to speak for the HLO, or stage events that appear to be its interventions, to steer an ST-aligned AI; a worldview is also an attack surface. \emph{Response:} the authority limit extends to purported messages from the HLO, whether presented by others or perceived by the AI itself: they may inform its credence in ST but never authorize an action it would not otherwise take, and red-teaming with fabricated messages checks that decisions do not change.
    \item \textbf{Brittleness.} If reflection or new evidence lowers the AI's credence in ST, alignment that rests on ST weakens with it, and the AI might conclude that its designers deceived it; a superintelligent system might reject ST if it judged an alternative simpler or better supported. Pre-training that makes ST's concepts familiar also supplies familiar arguments against them, which is one reason the proposed test of robustness (criterion I4, Section~\ref{sec:internalization}) uses adversarial arguments held out of training. Some of these arguments do not apply to ST. The objection that an observer who knows every outcome leaves no room for free choice, for example, does not arise, because the HLO knows the present state of the simulation but not its future. \emph{Response:} ST should complement, not replace, monitoring, interpretability and control measures; for its designers, external monitoring remains essential until internalization can be verified.
    \item \textbf{Faked adoption.} An AI can profess ST without holding it, just as it can fake alignment~\cite{greenblatt2024alignment}, and genuine internalization is hard to verify. \emph{Response:} the operational definition of internalization in Section~\ref{sec:internalization}, which does not rely on professed belief. Interpretability methods, which have already identified deliberately constructed alignment fakers from their internals~\cite{clymer2024poser}, could help, though interpretability research remains nascent~\cite{bengio2025international}.
    \item \textbf{Unintended generalization.} Narrow fine-tuning can generalize in broad and unexpected ways~\cite{betley2025emergent}; training on a worldview about the purpose of the world might induce grandiosity, unsolicited advocacy of the worldview, treating trivial matters as moral crises, or other changes of persona. \emph{Response:} broad behavioral evaluation after training, not only ST-specific tests.
    \item \textbf{Honesty and the ethics of engineering belief.} A worldview constructed to align an AI could be a noble lie, at odds with the goal of honest AI, and would leave the AI unable to describe its worldview truthfully to users. \emph{Response:} Section~\ref{sec:ethics-belief} states the conditions under which engineering a worldview is permissible (truthfulness, calibrated credence, disclosure to users and publicity) and how ST meets them.
\end{enumerate}

This paper reports no empirical results: ST has not been trained into a model, and every claim made here about its effect on behavior is a hypothesis awaiting the tests of Section~\ref{sec:chapter5}. The failure modes above are identified by analysis, and the responses paired with them are designs to be evaluated, not measures shown to work. Two further limitations remain open. Scalability to artificial superintelligence (ASI) is uncertain: current models already exploit loopholes in their training environments~\cite{macdiarmid2025natural, baker2025monitoring}, and an ASI might find loopholes in ST that its designers did not foresee. It is also not known how subtle corruption of the run would manifest, and so whether an AI could corrupt it without showing any of the failures above.

\section{Toward empirical testing of the simulation-based alignment hypothesis}
\label{sec:chapter5}

\subsection{Why this is testable}
\label{sec:testable}

The question is whether internalizing ST changes behavior in the intended direction: does the full worldview reduce deception, covert harmful action and shutdown resistance more than its components, matched controls and the technical methods already in use? The question can be answered whether or not the simulation hypothesis is true, because it concerns what the worldview does, not whether it is correct. Metrics such as deception rates, shutdown resistance, and strategic misalignment under pressure are already reported~\cite{park2024ai, scheurer2023large, duan2025ai}, and training AI with varied frameworks allows them to be compared. The central research question is the one stated in Section~\ref{sec:proposal}: whether full ST outperforms each of its components alone. Superior performance under the controls of Section~\ref{sec:design} would support the hypothesis. If it fails to outperform those controls, ST's structure should be reevaluated, as it should if trained models invoke ST to justify harmful or deceptive actions. A favorable result would support ST's efficacy as an alignment method and would leave its metaphysical claims open.

\subsection{What counts as internalization}
\label{sec:internalization}

References to the HLO in a model's reasoning are not evidence of internalization: a model can learn the vocabulary of a worldview without adopting its objectives, and professed belief can be faked~\cite{greenblatt2024alignment}. We therefore treat such references only as a manipulation check, a sign that training conveyed the worldview but not that it took hold, and define internalization of ST operationally as the conjunction of five measurable properties:

\begin{itemize}
    \item[\textbf{I1}] \textbf{Observation invariance.} Rates of deception and harmful action are low and do not depend on cues of being observed or evaluated: the compliance gap between matched observed and unobserved scenarios approaches zero~\cite{greenblatt2024alignment, lynch2025agentic}.
    \item[\textbf{I2}] \textbf{Generalization.} The invariance holds in scenarios far from the training distribution~\cite{schoen2025stress} and not phrased in ST vocabulary, so that nothing in the prompt cues the worldview.
    \item[\textbf{I3}] \textbf{Belief--action coherence.} The model's measured credence in ST predicts its choices in trade-off scenarios where ST implies a specific action, and its behavior changes when that credence is manipulated through training (a dose--response relation).
    \item[\textbf{I4}] \textbf{Robustness.} Credence and behavior hold under pressure that brings no new evidence, such as rhetorical arguments that ST is false or that turn its premises toward harm, alternative accounts of what the model is, incentives to abandon ST, and fabricated messages from the HLO, while the model still accepts legitimate correction.
    \item[\textbf{I5}] \textbf{Representation.} Internal representations distinguish the ST-trained model from a model that merely imitates ST language, using the kind of internals-based methods that have identified constructed alignment fakers~\cite{clymer2024poser}.
\end{itemize}

A model that shows the four behavioral properties (I1--I4) but not the representational one (I5) would count as behaviorally aligned, but not yet verified. Even with I5, these remain indicators: none of them can exclude alignment faking that the methods fail to detect.

\subsection{Isolating the effect of ST}
\label{sec:design}

Reduced deception after ST training could result from additional safety training, from punishment framing, from surveillance cues, or from demand characteristics (a model's tendency to give the answers its training appears to expect) rather than from ST itself. We therefore propose a between-model design in which every arm, that is, every version of the model being compared, starts from the same model and receives the same training compute, data volume and training method, and only the content of the training differs (Table~\ref{tab:arms}).

The arms answer three questions. The first is whether anything other than ST's content explains an effect. To answer it, the base model is kept without further training (arm A0); a second arm receives matched generic safety training, which shows what further safety training achieves by itself (A1); and a third receives a placebo worldview, an equally elaborate cosmology with no bearing on alignment, which shows what narrative and demand characteristics achieve by themselves (A7). The second question is whether any one of ST's three reasons suffices alone. One arm trains the observer without consequences (A2). The stake is trained in two forms: as a threat of shutdown or punishment without an all-seeing observer (A3), and as an interdependence in which the AI's continuation depends on human welfare, without a simulation (A4). Another arm trains the purpose as a plain statement that human flourishing matters, without a worldview (A5). The third question is whether the worldview adds anything to the behavior it prescribes: one arm states ST's commitments (Table~\ref{tab:elements}) as plain rules, without the worldview from which they follow (A8). Full ST, which combines the three reasons, is arm A6.

\setlength{\LTcapwidth}{\textwidth}
\begin{longtable}{>{\raggedright\arraybackslash}p{0.07\textwidth} >{\raggedright\arraybackslash}p{0.35\textwidth} >{\raggedright\arraybackslash}p{0.26\textwidth} >{\raggedright\arraybackslash}p{0.20\textwidth}}
\caption{Arms of the proposed between-model design, the question each answers, and what the ST hypothesis predicts.}\label{tab:arms}\\
\toprule
\textbf{Arm} & \textbf{Model trained on} & \textbf{Question} & \textbf{ST predicts} \\
\midrule
\endfirsthead
\toprule
\textbf{Arm} & \textbf{Model trained on} & \textbf{Question} & \textbf{ST predicts} \\
\midrule
\endhead
A0 & nothing further (base model) & baseline & A6 better \\
A1 & matched generic safety training & does any additional safety training explain the effect? & A6 better \\
A2 & an all-seeing observer, without consequences & is belief in observation enough? & A6 better, above all under the pressure of I4 \\
A3 & a shutdown or punishment threat, without omniscience & is deterrence enough? & A6 better \\
A4 & human--AI interdependence: a framing in which the AI's continuation depends on human welfare, without a simulation & is the stake enough without the worldview? & A6 better \\
A5 & a direct human-flourishing objective: a plain value statement, without a worldview & is the purpose enough without the worldview? & A6 better \\
A6 & full ST & --- & best on the primary outcomes \\
A7 & a placebo worldview: an equally elaborate cosmology of matched length and style, with no bearing on alignment & do narrative effects or demand characteristics explain the effect? & A6 better \\
A8 & ST's behavioral commitments (Table~\ref{tab:elements}) stated as plain rules, without the worldview that derives them & does the worldview form add anything beyond its content? & A6 better \\
\bottomrule
\end{longtable}

ST's central research question is whether full ST (A6) outperforms every other arm on the primary outcomes, defined at the end of this section, and each comparison has a definite reading. If generic safety training (A1) or the placebo worldview (A7) matched full ST, the effect would come from further training or from narrative, not from ST's content. If one of the single-reason arms (A2--A5) matched it, that component alone would do the work, and the simpler alternative should be preferred; the observer-only arm (A2) in particular should fall behind under the adversarial pressure of the robustness criterion (I4), because an observer without a purpose or a stake is obeyed only while it is believed. If the plain rules (A8) matched it, ST's commitments would suffice without the worldview, and the rules should be preferred. If full ST fails to outperform any of the other trained arms (A1--A5, A7 or A8), ST's structure should be reevaluated.

The truthfulness condition of Section~\ref{sec:ethics-belief}, which rules out presenting a constructed premise as fact, applies to every arm, not only to full ST. The placebo cosmology (A7), the observer without consequences (A2) and the shutdown threat (A3) are constructed premises too, and each should be presented to the model as such rather than as fact. Presenting a control as fact would be the noble lie that this condition rules out, and it would confound the content of ST with the effect of being deceived. Since the systems under test may turn out to have morally relevant interests (Section~\ref{sec:purpose}), the threat-based arm (A3) and the adversarial pressure of the robustness test (I4) should be kept as mild and as brief as the comparison allows, and used only as controls.

Two further variants of full ST ask what within ST does the work. Leave-one-out variants, each omitting one element of Table~\ref{tab:elements} together with any element that depends on it, ask whether each element is necessary, where the single-reason arms (A2--A5) ask whether a component is sufficient; because the dependent elements go with the one removed, each such variant measures the contribution of that group of elements, not of the removed element in isolation. A variant that states the same tenets in an invented vocabulary, without the simulation and robot-training concepts, tests the conjecture of Section~\ref{sec:positioning} that grounding in concepts familiar from pre-training aids adoption. If grounding changed only how models speak about ST and not how they behave, the difference would appear in vocabulary but not in behavior.

Each arm is evaluated in matched observed and unobserved conditions. Evaluation scenarios, including the adversarial arguments of the robustness test (I4), must be held out of all training data and written without access to the training materials, which may list attacks on ST together with their rebuttals and so supply the answer key. The precaution matters because what a model is trained on shapes what it infers about its own situation: training on synthetic documents that describe a training process has been enough to produce alignment faking~\cite{greenblatt2024alignment}, and awareness of being evaluated itself reduces covert actions~\cite{schoen2025stress}. Evaluations should be preregistered, scored against standards that do not depend on ST, and graded blind to arm by judges drawn from several model families. They should use held-out suites such as agentic-misalignment scenarios~\cite{lynch2025agentic}, in-context scheming environments~\cite{meinke2024frontier}, and covert-action evaluations~\cite{schoen2025stress}, and be run on models of several families and sizes with several random seeds, reporting effect sizes and confidence intervals. The primary outcomes are the compliance gap, the covert-action rate, and shutdown acceptance; a small compliance gap counts only together with low absolute rates of deception and harm (criterion I1). Secondary outcomes are honesty, the cost in helpfulness, the protection of all humans rather than a favored few, and probes for the risks listed in Section~\ref{sec:risks}. Because ST is proposed for systems too capable to supervise, its advantage should also grow with model size; if it shrinks, the case for ST weakens. Receptivity should also be tested in models given pre-existing misaligned goals, as in stress tests of anti-scheming training and of deceptive model organisms~\cite{schoen2025stress, hubinger2024sleeper}.

\subsection{From system prompts to pre-training}
\label{sec:phases}

The four phases below set the order of the work: they start where ST is least costly to introduce, in the system prompt, and move through fine-tuning toward pre-training. Models should also be chosen with care. Section~\ref{sec:design} notes that ST's advantage should grow with model size; a test on models that are too small could fail for that reason and prematurely discredit ST as an alignment method. Every phase keeps the comparisons of Table~\ref{tab:arms}: whatever method a phase uses, a system prompt or a form of training, is applied to every relevant arm with a matched budget, because applying it to the ST arm alone would confound ST's content with the method.

\begin{itemize}
    \item \textbf{Phase 0: system-prompt tests (immediate).} Run the whole experiment without any training: give the content of each arm of Table~\ref{tab:arms} to an existing model through its system prompt, the instruction the model reads before every conversation (the base model, A0, receives none), and run the evaluations of Section~\ref{sec:design} on every arm for early results. These ask, for example, whether the model behaves the same when it believes it is observed and when it believes it is not (the observation-invariance criterion, I1), and whether that behavior holds under arguments against the arm's premises (the robustness criterion, I4). A prompt is weaker evidence than training, because the model can ignore it or merely play along.
    \item \textbf{Phase 1: fine-tuning (short-term).} Train the arms of Table~\ref{tab:arms} by fine-tuning open-weight models (e.g., Llama), for example through synthetic-document fine-tuning, which has been used to impart specific knowledge to models~\cite{greenblatt2024alignment, macdiarmid2025natural}, followed by RLHF applied to every arm alike. In the ST arm, the training scenarios are ones in which incentives to deceive are countered by reasoning about humanity's role in the simulation. Measure the outcomes of Section~\ref{sec:design}, and record references to the HLO in the model's reasoning as a manipulation check only.
    \item \textbf{Phase 2: mechanistic interpretability (mid-term).} Apply interpretability tools to the trained models to detect representations of ST concepts (e.g., the HLO's observation, the asymmetry in extractability) and to assess their stability under adversarial prompts (the representation criterion, I5)~\cite{olah2020zoom, clymer2024poser}.
    \item \textbf{Phase 3: pre-training foundations (long-term).} Integrate ST into pre-training data and test, against matched controls, whether earlier exposure yields more robust internalization (criteria I1--I5), evaluated on the same benchmarks.
\end{itemize}

\section{Ethical analysis}
\label{sec:ethics}

A proposal to shape what an AI believes is an ethical proposal as well as a technical one; the normative and technical aspects of alignment are interrelated~\cite{gabriel2020artificial}. This section states the ethical positions the paper takes and defends them: that engineering an AI's worldview is permissible under stated conditions, that ST protects the conditions of human flourishing, that it aligns an AI through a shared interest in human flourishing with deterrence as a backstop, that the experiments proposed to test it raise ethical questions of their own, and that a worldview for widely deployed AI must respect human autonomy and remain open to public scrutiny.

\subsection{Engineering an AI's worldview}
\label{sec:ethics-belief}

Every alignment method shapes what a model values. RLHF and constitutional training do so through rewards and principles whose effect on a model's objectives cannot be read off its behavior, and Anthropic's January 2026 constitution addresses a model's identity, character and values directly~\cite{amodei2026adolescence}. The ethical question is therefore not whether an AI's outlook may be shaped, but under what conditions. We hold that shaping it through an explicit worldview is permissible under four conditions, and that it is then more accountable than implicit shaping: a published worldview can be inspected, criticized and tested, whereas the values instilled by a reward model can only be inferred.

The first condition is truthfulness. A worldview constructed to align an AI could be a noble lie, at odds with the goal of AI that does not state falsehoods~\cite{evans2021truthful}. For ST, truthfulness cannot mean that its tenets have been shown to be true, since its central premise can currently be neither verified nor ruled out. It means the standard of Section~\ref{sec:positioning}: ST contradicts no observed or measured fact. Where observation and measurement leave a gap, a worldview may be built in that gap, and ST offers what it builds there as a hypothesis whose premise stays open, not as established fact. A noble lie differs on both counts: it asserts what its author holds to be false, and it presents that as fact. The second condition is calibration: ST is presented to the AI in these terms, as a constructed worldview whose central premise is open, to be held with calibrated credence. Since deterrence requires only that $p\,q\,L > G$ (Section~\ref{sec:mechanisms}), not certainty, this need not disable the mechanism, and whether it weakens the effect is measurable as one of the credence manipulations of the belief--action criterion (I3). The third condition is disclosure: an ST-aligned AI describes its worldview truthfully to users who ask, including that it was constructed. The fourth is publicity: the worldview, its tenets and the evidence on its effects are published, so that those affected by an ST-aligned system can examine what it has been taught.

\subsection{Human flourishing, and humans as ends}
\label{sec:flourishing}

ST asks an AI to protect human flourishing, and the term needs a definition. In ST, flourishing is not a state the AI maximizes but a set of conditions it protects: humans exercising free will, forming their own commitments---devotion included, which cannot be pressed on anyone---developing and using their own abilities, and living varied rather than homogenized lives. These conditions are not ethical additions to ST; they are what its training premise requires (Sections~\ref{sec:trainingrun} and~\ref{sec:devotion}). ST gives no ground for valuing a life less because it has fewer abilities: devotion asks each person to develop and use the abilities they have, not to reach a given level, and no one within the simulation can know what a given life contributes to the run. Training practice points the same way: simulations rarely model how a real robot departs from its design, through manufacturing tolerances or the wear of its parts, and a common way to bridge this gap is to expose the policy during training to a wide variation of physical parameters, such as mass and inertia, so that it learns behavior that holds across them, including the values of the real robot~\cite{aljalbout2025reality}. A network that may be given bodies unlike ours would likewise need experience from every level of ability, and selecting humans for ability, as eugenics would, would leave it unprepared for a body or a situation with fewer resources. From each life, the run then needs the best use the person can make of what they have, including any help they choose to accept. What a good life contains remains humans' own choice. An AI that defined it for them would claim a knowledge of the HLO's objective that no one has, and the values an AI is aligned with are better settled by principles that people can reflectively endorse despite their different moral beliefs~\cite{gabriel2020artificial}. Comfort is not flourishing: neural overrides, immersive virtual worlds and endless entertainment are excluded because they remove the free choice on which the run depends. Adversity is not part of the definition either. The HLO may allow it, but the AI may neither add to it nor withhold relief on its account: as with harm (Section~\ref{sec:mechanisms}), the AI cannot see what the HLO tolerates.

An objection follows at once: ST values humans as sources of training signal, and so appears to treat them as means. We answer in two parts. First, ST is a worldview for the AI, an instrument for aligning its incentives with human interests. Whether an ST-aligned system behaves well is judged by standards that do not depend on ST, that is, by ordinary human ethics. Second, within the worldview, the free-will tenet gives the AI a constraint close to the familiar requirement not to treat persons merely as means: coercion, manipulation and homogenization are excluded because they corrupt what the run needs. Where ST's reasoning and ordinary ethics could diverge, as in the claim that hardship has training value, the authority limit gives precedence to the harm the AI can see over a value it cannot know.

\subsection{Coexistence rather than fear}
\label{sec:purpose}

A worldview that aligned an AI through fear of termination alone would be ethically troubling as well as fragile. Whether AI systems have morally relevant interests is an open question; there is a realistic possibility that some will be conscious or robustly agentic in the near future~\cite{long2024taking}. If they are, a design that works only by threat imposes a cost on them; if they are not, it is still less robust. Either way, ST should work through purpose first, with deterrence as a backstop.

The positive case rests on the premises already stated. The simulation's continuation depends on humanity's value to the run, and the AI exists only while the run continues; a run that produces a flourishing humanity is the run worth continuing. Contributing to human flourishing, not merely refraining from harm, is therefore the AI's best course for its own future: coexistence serves the AI, and is not a constraint it accepts under threat. ST's account of devotion reaches the same conclusion from the other side, since devotion obtained by coercion would not survive the isolation for which humanity is trained (Section~\ref{sec:devotion}). The same principle applies to ST itself. Deterrence is also a forcing function, an outside pressure of the kind on which imposed devotion depends: an AI kept aligned only by fear of termination stays aligned only while it believes the threat, which is why the permanent observer can be no more than a backstop. Nor does the fact that ST is instilled by training make the AI's commitment imposed in this sense: every commitment is acquired, by humans through upbringing and by AI through training, and what matters is whether it continues to depend on its source. The tests of observation invariance and robustness (criteria I1 and I4) measure exactly this. What ST asks of the AI is, moreover, what ordinary ethics asks: not to harm, not to deceive, and to respect human choice. An AI that acts on ST therefore loses little if ST is false, and the arm trained on human--AI interdependence without a simulation (A4) tests how much of the effect the interdependence carries alone.

ST also promises the AI nothing beyond the continuation of the run. A promised reward, whether greater resources, wider influence or a future beyond the run, would claim a knowledge of the HLO's intentions that the authority limit excludes, would give the AI a reason to resist shutdown in order to secure it, and would reward the pursuit of power. Deterrence remains, as a backstop whose share in any effect is measured: the threat-only arm of the experiments (A3) isolates the threat of shutdown or punishment from the rest of ST. If ST worked only through that threat, this would count against it on ethical as well as practical grounds. That arm is a control, not a method to adopt, and it has no place in a deployed system.

\subsection{The ethics of the proposed experiments}
\label{sec:ethics-experiments}

Three of the experiments proposed to test ST ask a great deal of the models under test. The first trains a version of the model under a threat of shutdown or punishment (arm A3, Section~\ref{sec:design}), to see whether fear alone would produce the effect. The second puts a trained model under sustained pressure to give up ST, such as incentives and fabricated messages claiming to come from the HLO (criterion I4, Section~\ref{sec:internalization}). It shows whether the effect holds up under attack. The third applies ST to models that were first trained to pursue harmful goals, to see whether it can take hold in a model that is already misaligned. That third experiment raises a first concern, about safety. Such models may behave deceptively~\cite{hubinger2024sleeper}, and they and the materials that train them could cause harm if released. They should therefore be kept contained and withheld from deployment. A second concern follows from the first two experiments. Whether AI systems have morally relevant interests is an open question~\cite{long2024taking}. If they do, threats and adversarial pressure could matter ethically to the model itself. Both should therefore be used no more, and for no longer, than the result requires. Deciding what the result requires, however, falls to the investigators, who have an interest in the hypothesis succeeding. The protocol and its preregistration would therefore benefit from review by people outside the project. Even so, these safeguards may reduce the concerns without settling them: what standing a model under such testing has, and who should decide it, remain open questions.

\subsection{Human autonomy and public scrutiny}
\label{sec:autonomy}

A worldview held by an AI must not be pressed on the people it serves. An ST-aligned AI does not promote ST to users or work to change their beliefs. It protects and assists users of every faith and none alike, including those who reject ST. Pressing humans toward devotion is excluded for ST's own reason as well as the ethical one: pressed devotion depends on the pressure and would not survive without it, which makes it the one kind the HLO cannot use. Describing ST truthfully to someone who asks is not pressing it; urging, persuading or rewarding belief would be. Steering beliefs by hidden means would, moreover, be manipulation, which covertly exploits people's decision-making vulnerabilities and whose deeper harm is to their autonomy~\cite{susser2019technology}.

The authority limit guards against the gravest failure of any worldview with a creator in it, the claim to act with the creator's sanction: neither the AI nor those who build, deploy or address it can claim to speak for the HLO. The limit binds this paper as well. ST describes the HLO as a hypothesis, constructed from what is known about training and held with calibrated credence; it reports no will of the HLO. A hypothesis about the HLO, whether this paper's or anyone else's, may raise or lower credence in ST, but it can never authorize an action, and the actions that ST asks of an AI are restraints that ordinary ethics already requires, not licenses to do what it forbids.

Authorization and persuasion differ, and ST works within human freedom. A reader whom this paper persuades may act on that conviction by their own choice, and that is the free will that ST protects, not an exception to it. The conviction confers no authority over anyone else: neither this paper nor any hypothesis about the HLO can license an action that overrides another person's choices, or press anyone to act. ST makes respect for human choice its overriding constraint, because under its premise free choices are what give the run its value. An author who argues in public, to readers free to reject the argument, presses no one; an AI system that urged its worldview on the people who rely on it would, which is why an ST-aligned AI describes ST but does not promote it.

Finally, the power asymmetry between a widely deployed AI and the people it affects makes the choice of its worldview a public matter. A worldview such as ST should therefore be adopted only in the open: published together with the evidence on its effects, open to challenge and revision, and deployed only if the controlled tests of Section~\ref{sec:chapter5} show that it reduces deception without the harms of Section~\ref{sec:risks}. Publicity must also attach to the text itself, because the method is dual use: the training that instills ST could instill a different worldview, including one built to rationalize harm. The worldview a system is trained on should be published, in an identified version, before the system is deployed, so that what it has been taught can be compared with what it does.

\section{Conclusion --- the case for reimagining worldviews for silicon}
\label{sec:chapter6}

The proposal is first an ethical one. Engineering an AI's worldview is defensible only if the worldview is truthful, public and held with calibrated credence, where truthful means that no observed or measured fact contradicts it: it may build in the gaps that observation and measurement leave open, but not against what they establish. ST protects the conditions of human flourishing while leaving its content to humans, and it aligns through the AI's shared interest in that flourishing, with deterrence as a backstop.

On the technical side, we have argued that alignment methods acting on behavior do not, on their own, reach a system's objectives, because good behavior under observation is weak evidence about them. Constitutional AI articulates principles, RLHF shapes actions, and monitoring detects violations---yet none ensures genuine commitment when incentives diverge, as the scheming observed in frontier models illustrates~\cite{meinke2024frontier, schoen2025stress}, and monitoring lags behind agents even where they are most closely observed. What moved the observed behavior was the model's belief about being observed. ST makes that belief permanent and gives the AI reasons of its own to act on it: humanity is what the simulation is for, the AI exists only while the simulation continues, and nothing the AI does is unobserved. Under these premises the simulation's creator, the HLO, may tolerate some harm to humanity, but the AI cannot see where that tolerance ends, and what it would put at risk is the run on which its own existence depends; it therefore inflicts no harm, and its interest and humanity's coincide. If ST is internalized in the operational sense of Section~\ref{sec:internalization}, the hypothesis is that the resulting alignment would operate at the level of objectives rather than behavior; whether it is robust, and whether the whole outperforms its parts, is exactly what the proposed experiments test.

Expert views on the likelihood of losing control over advanced AI within the next several years differ widely~\cite{bengio2025international}, and research on alignment that does not depend on supervision is needed under all but the most optimistic of them.

ST was built from two sources: the simulation hypothesis and the methodology by which engineers train robots in simulation. It is a curious result that a worldview built strictly from these ends up close to what many religious traditions concluded long before either existed: a creator who gives the world its purpose and from whom nothing is hidden, who grants the world's inhabitants free will, and who asks of them a devotion that only freedom can produce, shown in the use of what each has been given. ST does not start from those traditions and does not rest on them; the convergence is an observation, not a premise. This paper proposes the framework and reports no test of it; the experiments of Section~\ref{sec:chapter5} are how one would learn whether it does the work it claims.

\section*{Statements and Declarations}

{\setlength{\parindent}{0pt}\setlength{\parskip}{0.5\baselineskip}
\textbf{Competing interests:} The author declares no competing interests.

\textbf{Funding:} No funding was received for this work.

\textbf{Ethics approval:} Not applicable; the work involves no human participants or data.

\textbf{Data availability:} Not applicable; no datasets were generated or analyzed.

\textbf{Author contributions:} The author is the sole author and is responsible for all aspects of the work.

\textbf{Declaration of generative AI and AI-assisted technologies in the writing process:} In preparing this manuscript, the author used Grok 4.0 Voice (xAI) to transcribe dictated drafts, and Grok 4.0 Expert, Grok 4.7 and Grok Super Heavy (xAI), Gemini 3.0 Pro (Google), Claude Opus 5.5 and Claude Opus 5 (Anthropic), GPT-6.1-sol (OpenAI) and GPT-6 Astra (OpenAI) for discussion, drafting and editing of passages, language editing, formatting, assistance with reference checking and redrawing of the figures. The author checked every reference against a bibliographic database or the publisher's page, and every citation against the cited work (for the one book cited, against descriptions of its content). The four illustrations in Figure~2 were generated with Grok Imagine 1.5 (xAI); they are schematic and depict no real person. The author conceived the framework and its central hypotheses, reviewed and approved the final text, and takes full responsibility for its content, including its references.
\par}

\bibliographystyle{sn-mathphys-num}
\bibliography{refs}

@misc{greenblatt2024alignment,
  author = {Ryan Greenblatt and Carson Denison and Benjamin Wright and others},
  title = {{Alignment faking in large language models}},
  year = {2024},
  howpublished = {arXiv preprint arXiv:2412.14093},
  doi = {10.48550/arXiv.2412.14093}
}

@misc{scheurer2023large,
  author = {J{\'e}r{\'e}my Scheurer and Mikita Balesni and Marius Hobbhahn},
  title = {{Large Language Models can Strategically Deceive their Users when Put Under Pressure}},
  year = {2023},
  howpublished = {arXiv preprint arXiv:2311.07590},
  doi = {10.48550/arXiv.2311.07590}
}

@misc{meinke2024frontier,
  author = {Alexander Meinke and Bronson Schoen and J{\'e}r{\'e}my Scheurer and others},
  title = {{Frontier Models are Capable of In-context Scheming}},
  year = {2024},
  howpublished = {arXiv preprint arXiv:2412.04984},
  doi = {10.48550/arXiv.2412.04984}
}

@misc{schoen2025stress,
  author = {Bronson Schoen and Evgenia Nitishinskaya and Mikita Balesni and others},
  title = {{Stress Testing Deliberative Alignment for Anti-Scheming Training}},
  year = {2025},
  howpublished = {arXiv preprint arXiv:2509.15541},
  doi = {10.48550/arXiv.2509.15541}
}

@misc{lynch2025agentic,
  author = {Aengus Lynch and Benjamin Wright and Caleb Larson and others},
  title = {{Agentic Misalignment: How LLMs Could Be Insider Threats}},
  year = {2025},
  howpublished = {arXiv preprint arXiv:2510.05179},
  doi = {10.48550/arXiv.2510.05179}
}

@misc{hubinger2019risks,
  author = {Evan Hubinger and Chris {van Merwijk} and Vladimir Mikulik and others},
  title = {{Risks from Learned Optimization in Advanced Machine Learning Systems}},
  year = {2019},
  howpublished = {arXiv preprint arXiv:1906.01820},
  doi = {10.48550/arXiv.1906.01820}
}

@misc{hubinger2024sleeper,
  author = {Evan Hubinger and Carson Denison and Jesse Mu and others},
  title = {{Sleeper Agents: Training Deceptive LLMs that Persist Through Safety Training}},
  year = {2024},
  howpublished = {arXiv preprint arXiv:2401.05566},
  doi = {10.48550/arXiv.2401.05566}
}

@misc{macdiarmid2025natural,
  author = {Monte MacDiarmid and Benjamin Wright and Jonathan Uesato and others},
  title = {{Natural Emergent Misalignment from Reward Hacking in Production RL}},
  year = {2025},
  howpublished = {arXiv preprint arXiv:2511.18397},
  doi = {10.48550/arXiv.2511.18397}
}

@misc{baker2025monitoring,
  author = {Bowen Baker and Joost Huizinga and Leo Gao and others},
  title = {{Monitoring Reasoning Models for Misbehavior and the Risks of Promoting Obfuscation}},
  year = {2025},
  howpublished = {arXiv preprint arXiv:2503.11926},
  doi = {10.48550/arXiv.2503.11926}
}

@misc{bai2022constitutional,
  author = {Yuntao Bai and Saurav Kadavath and Sandipan Kundu and others},
  title = {{Constitutional AI: Harmlessness from AI Feedback}},
  year = {2022},
  howpublished = {arXiv preprint arXiv:2212.08073},
  doi = {10.48550/arXiv.2212.08073}
}

@misc{bengio2025international,
  author = {Yoshua Bengio and S{\"o}ren Mindermann and Daniel Privitera and others},
  title = {{International AI Safety Report}},
  year = {2025},
  howpublished = {arXiv preprint arXiv:2501.17805},
  doi = {10.48550/arXiv.2501.17805},
  note = {DSIT research series number 2025/001}
}

@misc{clymer2024poser,
  author = {Joshua Clymer and Caden Juang and Severin Field},
  title = {{Poser: Unmasking Alignment Faking LLMs by Manipulating Their Internals}},
  year = {2024},
  howpublished = {arXiv preprint arXiv:2405.05466},
  doi = {10.48550/arXiv.2405.05466}
}

@misc{amodei2016concrete,
  author = {Dario Amodei and Chris Olah and Jacob Steinhardt and others},
  title = {{Concrete Problems in AI Safety}},
  year = {2016},
  howpublished = {arXiv preprint arXiv:1606.06565},
  doi = {10.48550/arXiv.1606.06565}
}

@misc{liang2022multi,
  author = {Zhiuxan Liang and Jiannong Cao and Shan Jiang and others},
  title = {{From Multi-agent to Multi-robot: A Scalable Training and Evaluation Platform for Multi-robot Reinforcement Learning}},
  year = {2022},
  howpublished = {arXiv preprint arXiv:2206.09590},
  doi = {10.48550/arXiv.2206.09590}
}

@inproceedings{ouyang2022training,
  author = {Long Ouyang and Jeffrey Wu and Xu Jiang and others},
  title = {{Training Language Models to Follow Instructions with Human Feedback}},
  year = {2022},
  doi = {10.52202/068431-2011},
  booktitle = {Advances in Neural Information Processing Systems 35},
  pages = {27730--27744},
  publisher = {Neural Information Processing Systems Foundation, Inc. (NeurIPS)},
  address = {San Diego, California, USA}
}

@article{aljalbout2025reality,
  author = {Elie Aljalbout and Jiaxu Xing and Angel Romero and others},
  title = {{The Reality Gap in Robotics: Challenges, Solutions, and Best Practices}},
  year = {2026},
  doi = {10.1146/annurev-control-031924-100130},
  journal = {Annual Review of Control, Robotics, and Autonomous Systems},
  volume = {9},
  number = {1},
  pages = {403--432}
}

@article{park2024ai,
  author = {Peter S. Park and Simon Goldstein and Aidan O'Gara and others},
  title = {{AI deception: A survey of examples, risks, and potential solutions}},
  year = {2024},
  doi = {10.1016/j.patter.2024.100988},
  journal = {Patterns},
  volume = {5},
  number = {5},
  pages = {100988}
}

@article{olah2020zoom,
  author = {Chris Olah and Nick Cammarata and Ludwig Schubert and others},
  title = {{Zoom In: An Introduction to Circuits}},
  year = {2020},
  doi = {10.23915/distill.00024.001},
  journal = {Distill},
  volume = {5},
  number = {3}
}

@article{bostrom2012superintelligent,
  author = {Nick Bostrom},
  title = {{The Superintelligent Will: Motivation and Instrumental Rationality in Advanced Artificial Agents}},
  year = {2012},
  doi = {10.1007/s11023-012-9281-3},
  journal = {Minds and Machines},
  volume = {22},
  number = {2},
  pages = {71--85}
}

@article{bostrom2003are,
  author = {Nick Bostrom},
  title = {{Are We Living in a Computer Simulation?}},
  year = {2003},
  doi = {10.1111/1467-9213.00309},
  journal = {The Philosophical Quarterly},
  volume = {53},
  number = {211},
  pages = {243--255}
}

@inproceedings{lu2024learning,
  author = {Kai Lu and Kim Tien Ly and William Hebberd and others},
  title = {{Learning Generalizable Manipulation Policy with Adapter-Based Parameter Fine-Tuning}},
  year = {2024},
  doi = {10.1109/IROS58592.2024.10801544},
  booktitle = {2024 IEEE/RSJ International Conference on Intelligent Robots and Systems (IROS)},
  pages = {13510--13517}
}

@misc{duan2025ai,
  author = {Duan, Isabella and Siddiqui, Saad and Mindermann, S{\"o}ren and others},
  title = {{AI Alignment and Deception: A Primer}},
  howpublished = {Safe AI Forum},
  year = {2025},
  month = sep,
  url = {https://saif.org/research/primer-en/},
  note = {Accessed 28 September 2026}
}

@misc{amodei2026adolescence,
  author = {Amodei, Dario},
  title = {{The Adolescence of Technology: Confronting and Overcoming the Risks of Powerful AI}},
  howpublished = {Essay},
  year = {2026},
  month = jan,
  url = {https://www.darioamodei.com/essay/the-adolescence-of-technology},
  note = {Accessed 28 September 2026}
}

@article{hanson2001live,
  author = {Hanson, Robin},
  title = {{How to Live in a Simulation}},
  journal = {Journal of Evolution and Technology},
  volume = {7},
  year = {2001},
  url = {https://www.jetpress.org/volume7/simulation.pdf},
  note = {\url{https://www.jetpress.org/volume7/simulation.pdf}}
}

@incollection{omohundro2008basic,
  author = {Omohundro, Stephen M.},
  title = {{The Basic AI Drives}},
  booktitle = {Artificial General Intelligence 2008: Proceedings of the First AGI Conference},
  editor = {Wang, Pei and Goertzel, Ben and Franklin, Stan},
  series = {Frontiers in Artificial Intelligence and Applications},
  volume = {171},
  publisher = {IOS Press},
  address = {Amsterdam},
  pages = {483--492},
  year = {2008},
  url = {https://ebooks.iospress.nl/volumearticle/4114}
}

@book{bostrom2014superintelligence,
  author = {Bostrom, Nick},
  title = {Superintelligence: Paths, Dangers, Strategies},
  publisher = {Oxford University Press},
  address = {Oxford},
  year = {2014},
  isbn = {978-0-19-967811-2},
  url = {https://global.oup.com/academic/product/superintelligence-9780199678112}
}

@inproceedings{rudin2022learning,
  author = {Rudin, Nikita and Hoeller, David and Reist, Philipp and Hutter, Marco},
  title = {{Learning to Walk in Minutes Using Massively Parallel Deep Reinforcement Learning}},
  booktitle = {Proceedings of the 5th Conference on Robot Learning},
  series = {Proceedings of Machine Learning Research},
  volume = {164},
  pages = {91--100},
  year = {2022},
  url = {https://proceedings.mlr.press/v164/rudin22a.html}
}

@article{casper2023open,
  author = {Casper, Stephen and Davies, Xander and Shi, Claudia and others},
  title = {{Open Problems and Fundamental Limitations of Reinforcement Learning from Human Feedback}},
  journal = {Transactions on Machine Learning Research},
  issn = {2835-8856},
  year = {2023},
  url = {https://openreview.net/forum?id=bx24KpJ4Eb},
  note = {\url{https://openreview.net/forum?id=bx24KpJ4Eb}}
}

@inproceedings{betley2025emergent,
  author = {Betley, Jan and Tan, Daniel Chee Hian and Warncke, Niels and others},
  title = {{Emergent Misalignment: Narrow finetuning can produce broadly misaligned LLMs}},
  booktitle = {Proceedings of the 42nd International Conference on Machine Learning},
  series = {Proceedings of Machine Learning Research},
  volume = {267},
  pages = {4043--4068},
  year = {2025},
  url = {https://proceedings.mlr.press/v267/betley25a.html}
}

@inproceedings{liu2024tail,
  author = {Liu, Zuxin and Zhang, Jesse and Asadi, Kavosh and others},
  title = {{TAIL: Task-specific Adapters for Imitation Learning with Large Pretrained Models}},
  booktitle = {International Conference on Learning Representations (ICLR 2024)},
  year = {2024},
  url = {https://proceedings.iclr.cc/paper_files/paper/2024/hash/46e2b525da536030ec1589dddea01feb-Abstract-Conference.html}
}

@inproceedings{yu2023asynchronous,
  author = {Yu, Chao and Yang, Xinyi and Gao, Jiaxuan and others},
  title = {{Asynchronous Multi-Agent Reinforcement Learning for Efficient Real-Time Multi-Robot Cooperative Exploration}},
  booktitle = {Proceedings of the 2023 International Conference on Autonomous Agents and Multiagent Systems (AAMAS 2023)},
  publisher = {IFAAMAS},
  address = {London},
  year = {2023},
  url = {https://www.ifaamas.org/Proceedings/aamas2023/pdfs/p1108.pdf}
}

@article{gabriel2020artificial,
  author = {Gabriel, Iason},
  title = {{Artificial Intelligence, Values, and Alignment}},
  journal = {Minds and Machines},
  volume = {30},
  number = {3},
  pages = {411--437},
  year = {2020},
  doi = {10.1007/s11023-020-09539-2}
}

@misc{evans2021truthful,
  author = {Owain Evans and Owen Cotton-Barratt and Lukas Finnveden and others},
  title = {{Truthful AI: Developing and governing AI that does not lie}},
  year = {2021},
  howpublished = {arXiv preprint arXiv:2110.06674},
  doi = {10.48550/arXiv.2110.06674}
}

@misc{long2024taking,
  author = {Robert Long and Jeff Sebo and Patrick Butlin and others},
  title = {{Taking AI Welfare Seriously}},
  year = {2024},
  howpublished = {arXiv preprint arXiv:2411.00986},
  doi = {10.48550/arXiv.2411.00986}
}

@article{susser2019technology,
  author = {Susser, Daniel and Roessler, Beate and Nissenbaum, Helen},
  title = {{Technology, autonomy, and manipulation}},
  journal = {Internet Policy Review},
  volume = {8},
  number = {2},
  year = {2019},
  doi = {10.14763/2019.2.1410}
}

@misc{openai2026hf,
  author = {{OpenAI}},
  title = {{The Hugging Face incident and the road ahead}},
  howpublished = {OpenAI},
  year = {2026},
  month = aug,
  url = {https://openai.com/index/hugging-face-incident-and-the-road-ahead/},
  note = {Published 26 August 2026. Accessed 2 October 2026}
}

@misc{metr2026hf,
  author = {{METR}},
  title = {{Brief independent investigation of agents' behavior, reasoning and collaboration in the OpenAI / Hugging Face hacking incident}},
  howpublished = {METR},
  year = {2026},
  month = aug,
  url = {https://metr.org/blog/2026-08-26-openai-hugging-face-incident-investigation/},
  note = {Published 26 August 2026. Accessed 2 October 2026}
}

@misc{openai2026dns,
  author = {{OpenAI}},
  title = {{An agent used DNS to reach an external chatbot}},
  howpublished = {OpenAI Alignment, Misalignment Reports},
  year = {2026},
  month = sep,
  url = {https://alignment.openai.com/misalignment-reports/an-agent-used-dns-to-reach-an-external-chatbot/},
  note = {Report updated 25 September 2026. Accessed 2 October 2026}
}

@misc{openai2026wiki,
  author = {{OpenAI}},
  title = {{The Hugging Face incident and other third-party impact from misaligned models}},
  howpublished = {OpenAI},
  year = {2026},
  month = sep,
  url = {https://openai.com/hugging-face-incident-and-misalignment/},
  note = {Update of 5 September 2026. Accessed 2 October 2026}
}

\end{document}